\documentclass[doublecol]{epl2}
\usepackage{color}
\usepackage{amsmath}
\usepackage{amssymb}

\newcommand{\ignore}[1]{}

\title{Impact of individual nodes in Boolean network dynamics}

\author{Fakhteh Ghanbarnejad \and Konstantin Klemm}
\shortauthor{Ghanbarnejad \& Klemm}

\institute{
Bioinformatics, Institute for Computer Science,
Leipzig University, H\"{a}rtelstrasse 16-18, 04107 Leipzig, Germany}


\abstract{Boolean networks serve as discrete models of regulation and signaling
in biological cells. Identifying the key controllers of such processes is
important for their understanding and planning further analysis. We
quantify the dynamical impact of a node as the probability of damage spreading
after switching the node's state. The leading eigenvector of the adjacency
matrix is a good predictor of dynamical impact in case of long-term
spreading. Quality of prediction is further improved when eigenvector centrality
is based on the weighted matrix of activities rather than the unweighted
adjacency matrix. Simulations are performed with random Boolean networks and a
model of signaling in fibroblasts. The findings are supported by analytic
arguments from a linear approximation of damage spreading.}

\begin{document}

\maketitle
%%%%%%%%%%%%%%%%%%%%%%%%%%%%%%%%%%%%%%%%%%%%%%%%%%%%%%%%%%%%%%%%%%%%%%%%%%%%%%%%%%%

\section{Introduction}

%Boolean nets, general

Boolean networks are coarse-grained models of the regulatory dynamics that
controls the survival and proliferation of a living cell
\cite{Kauffman:1969,Kauffman:1993,deJong:2002,Bornholdt:2005}. The dynamics is
time- and state-discrete. This Boolean abstraction assumes that small
differences in concentration levels are irrelevant. The binary distinction
of a low or a high concentration of each bio-molecule is sufficient to
capture the dynamics. 

%ensembles of random BN

A purely theoretical branch of studies is devoted to randomly constructed Boolean
networks \cite{Kadanoff-review,drossel-review} and strives to elucidate generic
features of Boolean dynamics. From the perspective of statistical mechanics,
averaged macroscopic quantities in the limit of large system size are described
in dependence of ensemble parameters such as the probability distribution of the
employed Boolean functions \cite{Mihaljev:2006,Szejka:2008} and the degree
distributions of the networks \cite{Aldana:2003,Drossel:2009}.  The number of
attractors (ergodic subsets of the state space)
\cite{Kauffman:1993,Socolar:2003,Samuelsson:2003,Klemm:2005b} and the stability
under perturbations
\cite{Derrida:1986,Aldana:2003,Shmulevich:2003,Fretter:2009,Peixoto:2010,Schmal:2011,Mozeika:2011,Ghanbarnejad:2011}
have been investigated. The underlying fundamental result is a transition between
convergent (stable) and divergent (unstable) dynamics when the input sensitivity
of the Boolean functions passes a critical value
\cite{Derrida:1986,Seshadhri:2011}.

%specific networks for empirical systems

In recent years, the theory of random ensembles has been complemented by case
studies showing that suitably constructed Boolean networks capture the behaviour
of empirical regulatory systems
\cite{Kauffman:2003,Albert:2003,Li:2004,Helikar:2008,Albert:2008}. These
system-specific Boolean networks are obtained by compiling biochemical
interactions from the literature \cite{Davidich:2008}, by discretizing existing
models of differential equations \cite{Davidich:2008b}, or by inference from
data by a dedicated algorithm \cite{Xia:2011}.

With the advent of system-specific Boolean models, new conceptual questions and
analytical and numerical challenges arise. In particular, the response of the
system to external intervention may be quantified in a more detailed manner than
an averaging over all eligible perturbations. Since each node now represents a
specific biochemical entity, a node's individual impact on the dynamics is of
interest. The prediction of nodes' impacts from the model may be compared to
biological experiments. It is expected to trigger additional experiments and
lead to improvement of models.

The goal of this contribution is to establish a formal notion of node impact in
Boolean dynamics and its relation to a node's topological position in the
network. We perform a linear approximation of the long-term effect of a
perturbation at a specific node $i$. We find that, in good approximation, the
expected impact is monotonically related to the entry of $i$ in the leading
eigenvector of the adjacency matrix. When not only the network structure but
also the Boolean functions are known, the estimate is improved by replacing the
adjacency matrix with a weighted matrix of the activity values derived from the
functions. The analytic approximations are validated by numerical studies of
random Boolean networks and an empirical network from the literature.

%%%%%%%%%%%%%%%%%%%%%%%%%%%%%%%%%%%%%%%%%%%%%%%%%%%%%%%%%%%%%%%%%%%%%%%%%%%%%%
\section{Boolean networks}
%%%%%%%%%%%%%%%%%%%%%%%%%%%%%%%%%%%%%%%%%%%%%%%%%%%%%%%%%%%%%%%%%%%%%%%%%%%%%%

\begin{figure*}
\centerline{\includegraphics[clip,width=\textwidth]{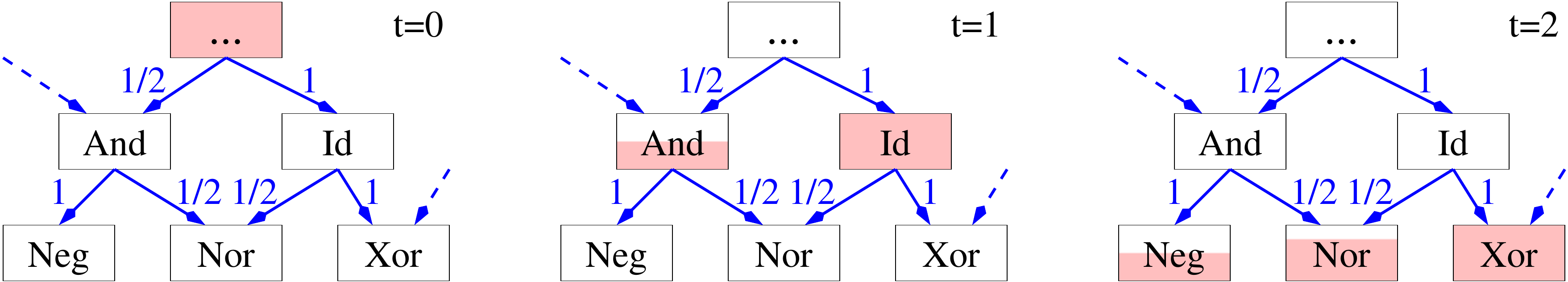}}
\caption{\label{fig:illu}
Probabilistic description of damage spreading in a Boolean network. The
estimated damage probability $p_i(t)$ for a node $i$ at time $t$ is indicated
by the height of the shaded area. At time $t=0$, the upper node is perturbed,
thus having a damage probability 1. Neglecting correlations, the probability
that a damage spreads from a node $i$ to a node $j$ is the
activity $\alpha_{ij}$ as a label on each connection $i \rightarrow j$.
Note that the case of more than one perturbed input, such as for the
node with the Nor-function, is not captured by the
activities. In the analytic treatment, we assume linear superposition
of damage probabilities. The node performing Nor has an
estimated damage probability $(1/2)(1/2)+1(1/2) =3/4$ at time $t=2$.
}
\end{figure*}

A Boolean network is a state- and time-discrete dynamical system. The dynamics
is defined by an iteration 
\begin{equation} \label{eq:dynamic_general}
x(t+1) = f(x(t))
\end{equation}
with $N$ Boolean dynamical variables written as a binary vector
$x(t) \in \{0,1\}^N$ at each time $t \in \mathbb{N}\cup\{0\}$. The
mapping $f:\{0,1\}^N \rightarrow \{0,1\}^N$ is typically sparse:
calculating the state $x_j(t+1)$ requires knowledge of the state $x_i(t)$ for
a few ($\ll N$) indices $i$ at the previous time step. When the system is
pictured as a directed network, the nodes
$\{1,2,\dots,N\}$ carry the dynamic variables
$x_1,x_2,\dots,x_N$ interacting along a relatively small number of directed
arcs. Subindices address components of a vector such that $x_j$
is the Boolean state of node $j$ and $f_j$ is its Boolean function.

In order to formalize and quantify these ideas, we
consider the $x_i$-dependence of $f$ as the mapping
\begin{equation}
\partial^{(i)} f_j (x) = \left\{\begin{array}{cl}
1 & \textrm{if } f_j (x) \neq f_j (x^{\updownarrow i})\\
0 & \textrm{otherwise}
\end{array}\right.
\end{equation}
This is the Boolean analogue of the usual partial derivative of a function, using
$x^{\updownarrow i}$ to denote state vector $x$ with its $i$-th
entry negated. Note that $\partial^{(i)} f$ also maps from
$\{0,1\}^N$ to $\{0,1\}^N$. By averaging $\partial^{(i)} f$ over all states with equal weight,
the activity of $i$ on $j$ is obtained as
\begin{equation}
\alpha_{ij}(f) = 2^{-N} \sum_{x \in \{0,1\}^N}  \partial^{(i)} f_j(x) 
\end{equation}
The activity $\alpha_{ij}(f)$ is the probability that a
perturbation (negation of state) at node $i$ causes a perturbation at node $j$
in the subsequent time step, assuming that all $2^N$ state vectors occur
with equal probability. The {\em sensitivity} of the Boolean function $f_i$ is the sum
of its incoming activities,
\begin{equation}
s_i(f) = \sum_{j=1}^N \alpha_{ji}~.
\end{equation}
Likewise, we define the {\em strength} of $f_i$ as the sum of outgoing activities
\begin{equation}
\sigma_i(f) = \sum_{j=1}^N \alpha_{ij}~.
\end{equation}

The directed network on the nodes $\{1,2,\dots,N\}$ obtained from $f$
contains an arc from node $i$ to node $j$ if and only if $\alpha_{ij}(f)\neq 0$.
The adjacency matrix $A$ of the network has an entry $a_{ij} = 1$ if
$\alpha_{ij}(f)\neq 0$ and $a_{ij}=0$ otherwise.

%%%%%%%%%%%%%%%%%%%%%%%%%%%%%%%%%%%%%%%%%%%%%%%%%%%%%%%%%%%%%%%%%%%%%%%%%%%%%%
\section{Dynamical impact}
%%%%%%%%%%%%%%%%%%%%%%%%%%%%%%%%%%%%%%%%%%%%%%%%%%%%%%%%%%%%%%%%%%%%%%%%%%%%%%

So far we have considered the average effect of a flip perturbation at the input $i$ 
of a Boolean function $f_j$ on the output. Now we ask about the long-term
behaviour of the whole system after a perturbation. We define
\begin{equation}
H_i(t) = \{x \in \{0,1\}^N : f^t(x) \neq f^t(x^{\updownarrow i}) \}
\end{equation}
as the set of initial conditions such that a perturbation at node 
$i$ spreads at least until time $t$. Then the fraction of such combinations
\begin{equation}
h_i(t) = \frac{|H_i(t)|} {2^N}
\end{equation}
out of all possible ones is the probability that the damage spreads for at least $t$
steps after perturbing node $i$. We call $h_i(t)$ the {\em dynamical impact}
of node $i$ for $t$ steps. Dynamical impact strongly varies across nodes
of a given network. For instance, the ratio between the largest and the average
impact is $20 \pm 4$ at $t=100$ on random Boolean networks with parameters
$N=500$, $K=2$ and $\langle s \rangle=1$.

Let us find an analytic approximation for $h_i(t)$ at long times $t$. By $p_i(t)$ we
denote the probability that node $i$ carries a damage at time $t$, i.e.\ the probability
that $[f^t(x)]_i \neq [f^t(x^{\updownarrow i})]_i$. After the perturbation has spread for
at least one time step, the damages and also the unperturbed states are correlated
across nodes in general. Then the single-node probabilities $p_i(t)$ are insufficient for an exact
description of the spreading probabilities.  Here we make an approximation by neglecting
the correlations. Then the damage probabilities follow the equation
\begin{equation}\label{eq:P}
p_j(t) \propto \sum_{i=1}^N \alpha_{ij} p_i(t-1)~.
\end{equation}
This equation is exact if the network, seen downstream from the initially perturbed node, is 
a directed tree. Then at most one term in the summation is non-zero.
Otherwise Eq.~(\ref{eq:P}) serves as an approximation assuming
a roughly linear accumulation of the damage.
Figure~\ref{fig:illu} provides an illustration.
In a more compact notation, Eq.~(\ref{eq:P}) reads
\begin{equation}
p(t) = \aleph^\textrm{T} p(t-1)
\end{equation}
using the transpose of the activity matrix $\aleph = (\alpha_{ij})_{ij}$. Iteration from the
initial condition yields
\begin{equation}
p(t) = (\aleph^\textrm{T})^t p(0)~.
\end{equation}
In the limit of large $t$, the projections on the (left and right) eigenspaces of the leading
eigenvector of $\aleph$ dominate the behaviour of $p$. Assuming that $\aleph$ is irreducible,
non-negativity ensures that these eigenspaces are one-dimensional by the Perron-Frobenius
theorem. Then we find unique normalized right and left principal eigenvectors 
$\epsilon^{\prime}$ and $\epsilon$ of $\aleph$ with non-negative entries.
In this approximation by the dominant eigenspaces, the evolution of $p$ reads
\begin{equation} \label{eq:dyadic}
p(t) = \lambda^t  (\epsilon^{\prime} \otimes \epsilon) p(0)
\end{equation}
with the dyadic product of $\epsilon$ and $\epsilon^\prime$ and the largest eigenvalue $\lambda$.
According to Equation~(\ref{eq:dyadic}), the projection
of the initial damage probability $p(0)$ on the eigenvector $\epsilon$ is what determines the
expected damage at long time $t$. In other words, $\epsilon_i$ is indicative of the long-term
damage expected from a perturbation at node $i$ in the linearized treatment
with suppression of correlations.

To which extent does this asymptotically expected damage amplitude $\epsilon_i$
inform us about the probability $h_i(t)$ that the perturbation spreads for a
long time $t$? In the following sections we investigate the question by
simulations. Often the network structure is known but information on the
Boolean functions lacking. Taking all non-zero activities as having value $1$
turns the
activity matrix into the adjacency matrix. Therefore we also consider the
predictive power of the leading left eigenvector $e=(e_1,e_2,\dots,e_N)$ of the
adjacency matrix. In situations without global knowledge on the system,
we may want to compare dynamical impacts of a few nodes, for which the
network neighbourhood is known. Then we can resort to the
strength $\sigma_i$ or the out-degree $d_i$ as centralities of node $i$
based on local information. Table~\ref{tab:measures} summarizes the four node
centralities under consideration.

\begin{table}
\caption{\label{tab:measures} Centrality measures considered as predictors for the dynamical impact $h_i(t)$.}
\centering
\smallskip

\begin{tabular}{|c|l|l|}
 \hline
$\downarrow$ range $\downarrow$  & adjacency matrix $A$ & activity matrix $\aleph$  \\ \hline 
local              & out-degree ($d_i$)                & strength ($\sigma_i$)      \\ \hline
global             & eigenvector  ($e_i$)    & eigenvector ($\epsilon_i$)           \\  \hline
\end{tabular}
\end{table}

%%%%%%%%%%%%%%%%%%%%%%%%%%%%%%%%%%%%%%%%%%%%%%%%%%%%%%%%%%%%%%%%%%%%%%%%%%%%
\section{Results for random networks}
%%%%%%%%%%%%%%%%%%%%%%%%%%%%%%%%%%%%%%%%%%%%%%%%%%%%%%%%%%%%%%%%%%%%%%%%%%%%

\begin{figure*}
\centerline{\includegraphics[clip,width=0.83\textwidth]{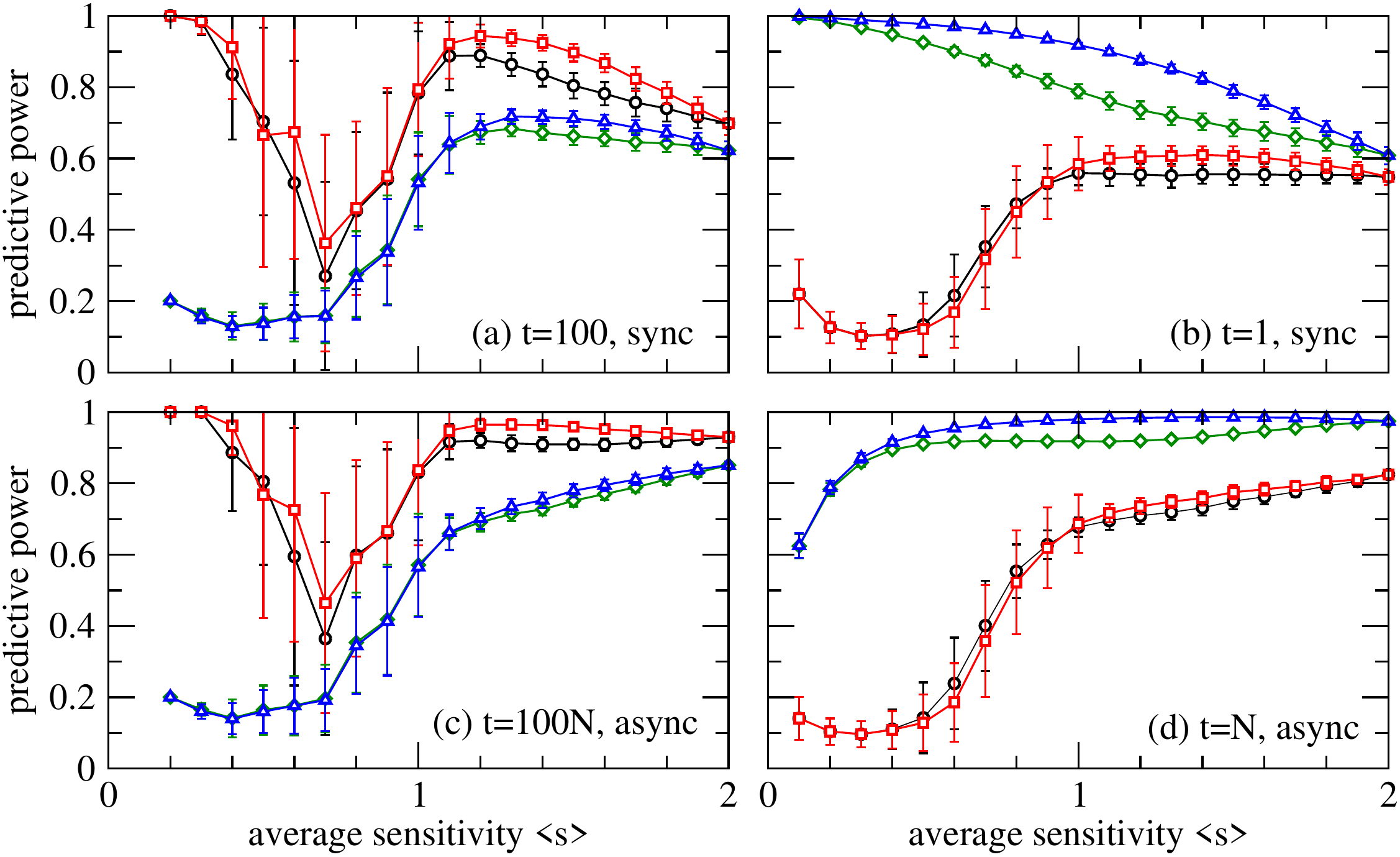}}
\caption{\label{fig:tmax}
Quality of prediction of dynamical impact in random Boolean networks at varying
average sensitivity $\langle s\rangle$. Symbols distinguish the centrality
measures out-degree $d$ (green $\diamond$), strength $\sigma$ (blue $\triangle$), and
the principal eigenvectors $\epsilon$ and $e$ of the activity matrix (red $\Box$)
and the adjacency matrix (black $\circ$). The four panels represent combinations of
long- or short-term prediction with deterministic synchronous or stochastic
asynchronous update. Each data point gives the rank order correlation
(cf.\ Methods) with dynamical impact $h(t)$,
averaged over 100 independent realizations of random Boolean network with given
sensitivity $\langle s \rangle$, $K=2$, and $N=500$ nodes.  The error bars
indicate the standard deviation over realizations. 
}
\end{figure*}

Let us investigate the dynamical impact of nodes  and its prediction by
centrality measures (cf.\ Table~\ref{tab:measures}) on random Boolean networks
with $N=500$ nodes and connectivity parameter $K=2$. See
Methods for details. As shown in Figure~\ref{fig:tmax}(a), the
long-term impact of perturbations is best predicted by the leading eigenvector
$\epsilon$ of the activity matrix in the whole range of sensitivity. Prediction
by the leading eigenvector $e$ of the adjacency  matrix is inferior to that by
$\epsilon$ in the supercritical regime $\langle s \rangle >1$. When reaching
$\langle s \rangle = K = 2$, predictive powers become equal again, because all
Boolean functions are exclusive-Or or its negation. Then all network connections
have activity value 1  and adjacency and activity matrices are the same. The
superiority of the eigenvector $\epsilon$ as a predictor is in agreement with
the analytic arguments given in the previous section. Slightly above the
critical sensitivity value 1, predictive  power shows a peak for all centrality
measures considered. Further analyses of the dynamics are necessary to
understand the variation of the predictive power with average sensitivity,
especially the minimum of $\mathcal{P}_\epsilon$ at $\langle s \rangle \approx
0.7$.

Figure~\ref{fig:tmax}(b) displays predictive power at short times, here $t=1$.
As expected, strength $\sigma$ is the best predictor in this case.
Predictions by the out-degree vector $d$ perform second best but significantly
worse than those by strength $\sigma$. 

The results in the upper panels of Fig.~\ref{fig:tmax} are obtained under
synchronous update of the whole system, as defined by
Eq.~(\ref{eq:dynamic_general}). In order to check the robustness of the results,
we repeat the simulations under stochastic {\em asynchronous} update according to
Equation~(\ref{eq:async}). The results, shown in panels (c) and (d) of
Fig.~\ref{fig:tmax}, are qualitatively similar to those obtained under
synchronous update. In the supercritical regime, however, the predictive power
of all four centrality measures is increased when the updating is asynchronous
instead of synchronous. Thus damage spreading is easier to predict under
asynchronous update, at least with the four centrality measures studied here.
This effect must be rooted in the interplay between the update order and the
network structure. For instance, the damage definitely heals when the perturbed
node receives the first update before all its predecessors. The frequency of
this happening decreases with the out-degree $d_i$ and incurs an additional
dependence of dynamical impact on the centrality measure $d$.

Simulations at different network sizes ($N=50$, $N=100$, not displayed)
yield similar results for all four combinations of long- or short-term
spreading and synchronous or asynchronous updates. The predictive power of all
four centrality measures remains constant or increases with system size.
Furthermore, we investigate networks with positive feedback
only, i.e.\ without negation of signals (see Methods). For $N=500$,
long-term prediction ($t=N$) and synchronous update, the
predictive power $\mathcal{P}_\epsilon$ of the eigenvector
is on average $0.80 \pm 0.18$; that of strength ($\mathcal{P}_\sigma$)
is on average $0.63 \pm 0.14$.

%%%%%%%%%%%%%%%%%%%%%%%%%%%%%%%%%%%%%%%%%%%%%%%%%%%%%%%%%%%%%%%%%%%%%%%%%%%%
\section{Switching between attractors}
%%%%%%%%%%%%%%%%%%%%%%%%%%%%%%%%%%%%%%%%%%%%%%%%%%%%%%%%%%%%%%%%%%%%%%%%%%%%

\begin{figure}
\centerline{\includegraphics[clip,width=0.42\textwidth]{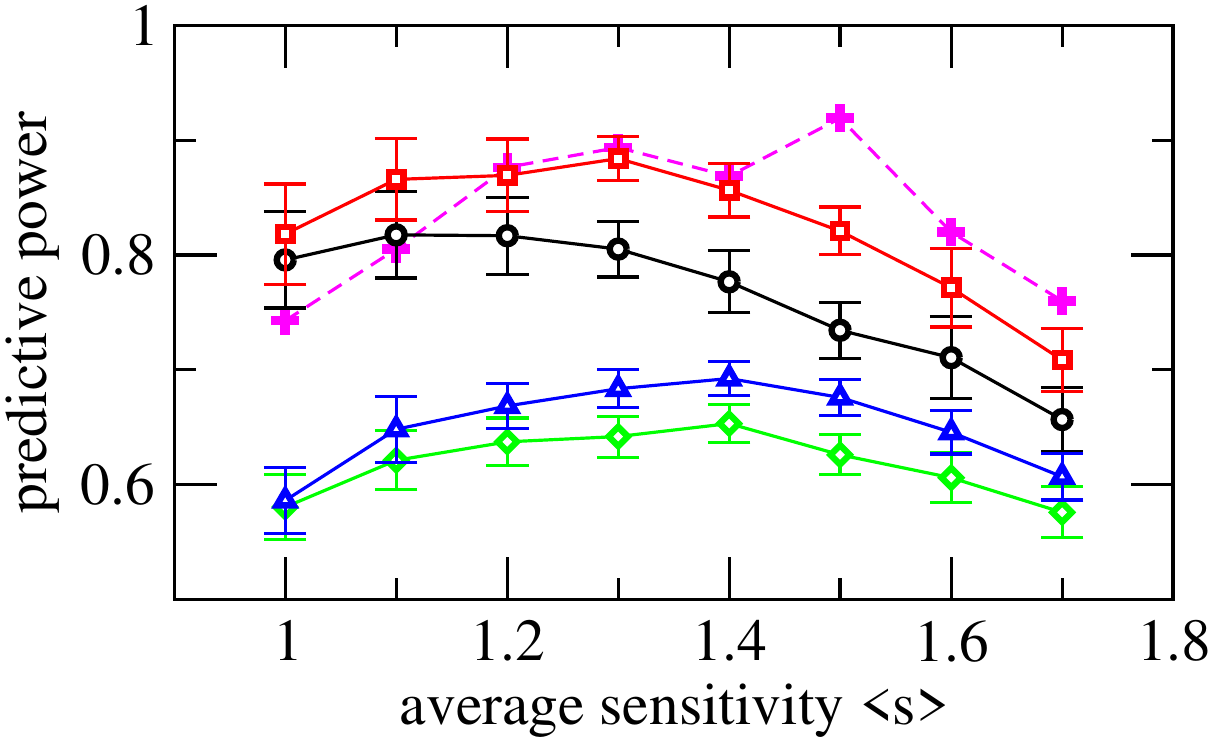}}
\caption{\label{fig:attr}
Power of centrality measures for predicting if a perturbation
changes the attractor reached. Symbols for the
centrality measures are the same as in Fig.~\protect\ref{fig:tmax}.
Each data point is an average over 100 random Boolean networks with $N=50$ nodes and
connectivity parameter $K=2$.
Error bars indicate the standard deviation over the random network ensemble.
Error bars are scaled down by the factor 0.2 to avoid overlapping.
Additional points ($+$-symbols with dashed lines) are the fraction of
realizations for which the principal eigenvector $\epsilon$ gives the best prediction
of all four measures.}
\end{figure}

The long-term behaviour of Boolean dynamics is determined by attractors. These
are minimal ergodic sets in state space. Under synchronous update, an attractor
of length $l$ is a sequence of states $x(0), x(1), \dots, x(l-1)$ such that
$f(x(t))= x([t+1]\bmod l)$ for all $t \in \{0,\dots,l-1\}$.

It is natural to ask if a perturbation in the initial condition will cause the
system to arrive at a different attractor. For this investigation, we define the
attractor impact $h_i^\prime$ of node $i$ as the fraction of initial conditions
where a perturbation at node $i$ changes the attractor eventually reached. At
difference with dynamical impact $h_i(t)$, attractor impact  $h_i^\prime$ does
not set an explicit time $t$ after which to determine the spreading or healing
of the perturbation. On the other hand, $h_i^{\prime}$ does count the
perturbation as healed whenever the perturbed and unperturbed dynamics
eventually become equal up to a time lag.

Figure~\ref{fig:attr} shows the predictive power of the centrality measures for
attractor impact of nodes. For averaged values, the
performance comparison yields
$\mathcal{P}_\epsilon > \mathcal{P}_e >  \mathcal{P}_\sigma >
\mathcal{P}_d$, being the same ordering as for predicting long-term dynamical impact.
Due to fluctuations around the averages, this ordering does not hold
for each single realization. At each considered value of the average sensitivity, a 
fraction at least $3/4$ of the realizations has $\epsilon$ as the best predictor. Subcritical networks,
$\langle s \rangle < 1$, are disregarded because here most realizations
do not have more than one attractor. For $\langle s \rangle > 1.7$,
attractor search exceeds available computer time.

%%%%%%%%%%%%%%%%%%%%%%%%%%%%%%%%%%%%%%%%%%%%%%%%%%%%%%%%%%%%%%%%%%%%%%%%%%%%
\section{Dynamical impact in a real network}
%%%%%%%%%%%%%%%%%%%%%%%%%%%%%%%%%%%%%%%%%%%%%%%%%%%%%%%%%%%%%%%%%%%%%%%%%%%%

\begin{table}
\caption{\label{tab:empirical}
Predictive power of centrality measures for the fibroblast signal transduction dynamics.
The upper part of the table considers the original system. The lower
part is for the system after removal of the nine nodes providing constant input. 
Each line of the table is a scenario defined by the update mode and the choice of
long- or short-term dynamics. The bold number indicates the maximum in each line.
}
\centering
\smallskip
\begin{tabular}{cc|c|c|c|c|l}
\cline{3-6}
& & \multicolumn{4}{|c|}{all nodes} \\ \cline{3-6}
& & $\mathcal{P}_{\epsilon}$ & $\mathcal{P}_e$ & $\mathcal{P}_{\sigma}$ & $\mathcal{P}_{d}$ \\ \cline{1-6}
\multicolumn{1}{|c|}{sync} &
\multicolumn{1}{|l|}{$t=1$} & 0.671	&0.454	&{\bf 0.930}	&0.455 &     \\ \cline{2-6}
\multicolumn{1}{|c|}{}                        &
\multicolumn{1}{|l|}{$t=100$} & {\bf0.920} &0.734	& 0.746	&0.523 &     \\ \cline{1-6}
\multicolumn{1}{|c|}{async} &
\multicolumn{1}{|l|}{$t=N$} & 0.706	&0.528	&{\bf 0.904} &0.564  \\ \cline{2-6}
\multicolumn{1}{|c|}{}                        &
\multicolumn{1}{|l|}{$t=100N$} & {\bf 0.854} &0.694	&0.748	&0.542  \\ \cline{1-6}
\end{tabular}
\begin{tabular}{cc|c|c|c|c|l}
\cline{3-6}
& & \multicolumn{4}{|c|}{only core nodes} \\ \cline{3-6}
& & $\mathcal{P}_{\epsilon}$ & $\mathcal{P}_e$ & $\mathcal{P}_{\sigma}$ & $\mathcal{P}_{d}$ \\ \cline{1-6}
\multicolumn{1}{|c|}{sync} &
\multicolumn{1}{|l|}{$t=1$} &0.633	&0.467	&{\bf 0.946} &0.528&     \\ \cline{2-6}
\multicolumn{1}{|c|}{}                        &
\multicolumn{1}{|l|}{$t=100$} &{\bf 0.911} &0.777	&0.738	&0.611&     \\ \cline{1-6}
\multicolumn{1}{|c|}{async} &
\multicolumn{1}{|l|}{$t=N$} &0.658	&0.543	&{\bf 0.919} &0.656&  \\ \cline{2-6}
\multicolumn{1}{|c|}{}                        &
\multicolumn{1}{|l|}{$t=100N$} &{\bf 0.834} &0.731	&0.741	&0.631&  \\ \cline{1-6}
\end{tabular}
\end{table}

\begin{table}
\caption{\label{tab:ranks}
The five core nodes of the fibroblast network with the largest dynamical impact
and their ranks with respect to the four centrality measures. Synchronous update
is performed on the core of the network, after removal of the nine input nodes.
Dynamical impact $h_i(t)$ measures spreading over $t=100$ time steps.
}
\centering
\smallskip

\begin{tabular}{|c|r|r|r|r|r|} \hline
node $i$   &$h_i(100)$&$r_i(\epsilon)$   &    $r_i(e)$  & $r_i(\sigma)$ & $r_i(d)$\\ \hline \hline
Src             &0.7707   &1      &1      &1      &1\\
B-Arrestin      &0.7061   &4      &4      &9      &14\\
GRK             &0.6458   &16     &27     &17     &43\\
PIP2-45         &0.5961   &2      &12     &4      &4\\
PKC             &0.5910   &3      &13     &3      &5\\ \hline
\end{tabular}
\end{table}  

Let us test the performance of predictors on a non-random network now. Helikar
et al.\ describe signal transduction in fibroblasts with a detailed Boolean
network \cite{Helikar:2008,Rue:2010}. The network has $N=139$ nodes and $548$
connections, including $59$ self-couplings. We choose this network
because of its size and because of its
large number of intertwined feedback loops of various lengths $l$, see also Figure 1 in \cite{Rue:2010}.
We quantify the abundance of feedback by the trace of the $l$-th power of the adjacency matrix $A$, finding
$\mathop{\rm tr}(A) =59$, $\mathop{\rm tr}(A^2)/2=568$, $\mathop{\rm tr}(A^3)/3=82455$ and
$\mathop{\rm tr}(A^4)/4=13921796$. The nodes fall into two classes. There
are 9 input nodes with a self-coupling. Each of these applies the identity
function to its own state, not receiving signals from any other node. These
nodes provide constant but choosable input to the rest of the network. Each of
the remaining 130 nodes receives an input from at least one other node in this
set. We call these the {\em core nodes}. The in-degree of nodes varies from 1 to
14, the out-degree varies from 1 to 28.

In table~\ref{tab:empirical}, we summarize the predictive power of centrality
measures for dynamical impact of nodes in the fibroblast network. Also for this
network, the leading eigenvector $\epsilon$ of the activity matrix is the
best predictor of a node's ability to cause long-term spreading of a perturbation.
Short-term spreading is best predicted by a node's strength $\sigma_i$. 
Table~\ref{tab:ranks} shows the five nodes with the largest dynamical impact and their
ranks with respect to the centrality measures. Prediction of these ranks by the centrality
measures  is not perfect. However, the leading eigenvector $\epsilon$ of the activity matrix
correctly identifies four out of the five nodes with the largest impact.

Table~\ref{tab:ranks} and the lower part of Table~\ref{tab:empirical} are
obtained for the fibroblast network after removal of the nine input nodes. These
nodes indefinitely sustain their state. Therefore a perturbation at an input
node $i$ never heals, yielding  maximal dynamical impact $h_i(t)=1$ for all
times $t$. Reduction to the dynamical core by the removal of the input nodes
allows for a less biased assessment of predictive power.

%%%%%%%%%%%%%%%%%%%%%%%%%%%%%%%%%%%%%%%%%%%%%%%%%%%%%%%%%%%%%%%%%%%%%%%%%%%%
\section{Discussion}
%%%%%%%%%%%%%%%%%%%%%%%%%%%%%%%%%%%%%%%%%%%%%%%%%%%%%%%%%%%%%%%%%%%%%%%%%%%%

Even in random Boolean networks, nodes exhibit significant differences in
dynamical impact. These differences are captured well by
local and global centrality measures. From a linearization of the
dynamics, these centralities arise as column sums and eigenvectors of
a network matrix. Practical applications therefore  benefit from efficient
computation as compared to costly direct simulation of damage spreading.
For the fibroblast network, a modern workstation calculates the
principal eigenvectors $\epsilon$ and $e$ in $<10^{-2}$\ s, to be
compared to several minutes for sampling over perturbations. Detection of
attractor switching with $N=500$ nodes takes time on the
order of hours, even days for some of the instances.

An important implication for networked systems is the possibility of
capturing response to perturbations based on incomplete information about the
system's structure. Here, prediction of dynamical impact only uses the activity
matrix while being ignorant of the actual rule tables. 
No distinction between positive and negative feedback enters the calculation. 
Comparing the case of randomly mixed feedback types to that
of only one type, we find that a combination of feedback
types is neither necessary nor detrimental for prediction of dynamical impact.

The scenario of predicting node impact based on partial knowledge is
particularly relevant for biological systems where not all interactions are
known in full detail. A large number of measures has been suggested for
identifying the dynamical centers of biological systems based on the underlying
network structure alone \cite{Wuchty:2003,Kitsak:2010,Klemm:2011}. Most
of these approaches provide only an intuitive understanding of the assumed
correlation between the centrality in the network and impact on the dynamics.
The present framework, beside its accuracy and computational efficiency, is
based on a verifiable description of the system's response to perturbations.
In particular, it allows to distinguish between short- and long-term effects.
The result is a detailed set of predictions testable in experiments.
In how far the predictions meet experimental outcomes depends on
the validity of idealizations at two levels: (i) the approximate analysis of
dynamical impact in this letter; (ii) the Boolean idealization to capture
the real system's dynamics.

%%%%%%%%%%%%%%%%%%%%%%%%%%%%%%%%%%%%%%%%%%%%%%%%%%%%%%%%%%%%%%%%%%%%%%%%%%%%
\section{Methods} \label{sec:methods}
%%%%%%%%%%%%%%%%%%%%%%%%%%%%%%%%%%%%%%%%%%%%%%%%%%%%%%%%%%%%%%%%%%%%%%%%%%%%

A random instance of a Boolean network with
$N$ nodes, connectivity parameter $K$ and expected average sensitivity $\langle
s \rangle$ is generated as follows. Each node $i$ is assigned a Boolean function
$f_i$, drawn from the distribution $\pi(f) \propto \exp[\lambda s(f)]$.
This distribution is normalized and supported by the set of
$2^K$ Boolean functions with at most $K$ inputs.
Here $\lambda$ is chosen such that the expectation value of $s(f)$ under
the distribution $\pi$ is equal to the
average sensitivity $\langle s \rangle$ \cite{Ghanbarnejad:2011}.
Then $\pi$ is the unique distribution maximizing entropy with the given 
$\langle s \rangle$. For each input, on which $f_i$ actually depends,
a link $(j,i)$ is established with the source node $j$ drawn uniformly at
random. When this would lead to a duplicate or self-coupling, $j$ is discarded
and redrawn. For the random networks with positive feedback only,
we use $\pi(f)=0.5$ if $f$ is the {\sc AND} or the {\sc OR} function,
$\pi(f)=0$ otherwise.

Both for the random and the empirical Boolean networks, we estimate 
dynamical impact of a node $i$ by $10^4$ runs of the dynamics.
For each of these, a state $x(0) \in \{0,1\}^N$ is drawn uniformly. Then
two replica of the system are initialized with $x(0)$ and at
$(x(0))^{\updownarrow i}$. The fraction of runs where the replica
are in different states at time $t$ is taken as approximation of $h_i(t)$.
When $h_i(t)$ is the same for all nodes $i$ or the largest eigenvalue of the
network's activity matrix is degenerate, the network is discarded and
a new independent realization is drawn. Discarding of network happens
mostly at small $\langle s \rangle$. It does not occur in any of the trials
with $\langle s \rangle \ge 1.2$~. 

The dynamics of Equation~(\ref{eq:dynamic_general}) is deterministic with synchronous 
update. Alternatively, we consider stochastic asynchronous update
as follows. At each time step $t$, a node $u(t)$ is drawn uniformly at random
and the nodes take states  
\begin{equation} \label{eq:async}
x_i(t+1) = \left\{ \begin{array}{rl}
f_i(x(t)) & \textrm{if } i=u\\
x_i(t)    & \textrm{otherwise}
\end{array} \right.
\end{equation}
in the subsequent time step. The same random sequence $u(t)$ of updated nodes is
used for the perturbed and the unperturbed replica of the system.

We quantify the predictive power $\mathcal{P}_y$ of a centrality measure
$y \in \{d,e,\sigma,\epsilon\}$ as the rank order correlation with
dynamical impact $\mathcal{P}_y = \mathop{\textrm{corr}} (r(h),r(y))$
using the usual Pearson correlation coefficient $\mathop{\textrm{corr}}$.
For a general vector $v=(v_1,v_2,\dots,v_n)$, the rank vector $r(v)$
has entries
\begin{equation}
r_i(v) =
1 + | \{ j \neq i | v_j > v_i \}| + \frac{1}{2}|\{j \neq i | v_j = v_i \}|~.
\end{equation}

%%%%%%%%%%%%%%%%%%%%%%%%%%%%%%%%%%%%%%%%%%%%%%%%%%%%%%%%%%%%%%%%%%%%%%%%%%%%
\section{Acknowledgements}
%%%%%%%%%%%%%%%%%%%%%%%%%%%%%%%%%%%%%%%%%%%%%%%%%%%%%%%%%%%%%%%%%%%%%%%%%%%%
The authors thank Thomas Skodawessely for a critical reading of the draft.
This work was supported financially by Volkswagen\-Stiftung.

\bibliographystyle{eplbib}
\bibliography{paper02}

\end{document}